\documentclass[twocolumn]{article}

\usepackage[bookmarks=false]{hyperref}

\usepackage{fullpage,psfrag,amsmath,amssymb,amsfonts,verbatim,mathtools,enumerate}
\usepackage[nolist]{acronym}
\usepackage[font=scriptsize,belowskip=-10pt]{caption}
\usepackage{subcaption}

\usepackage{graphicx}
\graphicspath{ {./} }

\newcommand{\cccc}[1]{ }

\newcommand{\reals}{{\mbox{\bf R}}}

\newcommand{\argmin}{\mathop{\rm argmin}}

\newcommand{\eyem}{I}
\newcommand{\vsn}{\Upsilon} % Graph of virtual sensor network.
\newcommand{\vsens}{V} % Set of virtual sensors.
\newcommand{\vnum}{g} % Number of virtual sensors.
\newcommand{\vlinks}{E} % Set of virtual links.
\newcommand{\vdemand}[1]{R(#1)} % Virtual sensor required capacity.
\newcommand{\tindx}{k} % Discrete time index.
\newcommand{\vsncenter}{c} % Virtual sensor network geographical center.
\newcommand{\vsndiam}{\delta} % Virtual sensor network geographical diameter.
\newcommand{\obs}{x} % Observation vector.
\newcommand{\obssize}{M} % Number of observation.
\newcommand{\param}{\theta} % Parameters vector.
\newcommand{\tparam}{\param} % Target Parameters vector.
\newcommand{\paramsize}{N} % Number of parameters to estimate.
\newcommand{\noise}{u} % Noise vector.
\newcommand{\model}{H} % Linear model matrix.
\newcommand{\aux}{z} % auxilary variable
\newcommand{\lagr}{\mathcal{L}} % Lagrangian.
\newcommand{\lagrmul}{\lambda} % Lagrangian multiplier.
\newcommand{\lagrpen}{\rho} % Lagrangian penalty parameter.
\newcommand{\psn}{\mathcal{G}} % Graph of substrate sensor network.
\newcommand{\psens}{S} % Set of physical sensors.
\newcommand{\assign}{A} % Set of physical sensors.
\renewcommand{\path}{P} % Set of physical sensors.
\newcommand{\mapf}{{\mathcal{M}_{A}}} % Set of physical sensors.
\newcommand{\pnode}{s} % A physical sensor.
\newcommand{\pnum}{n} % Number of physical sensors.
\newcommand{\plinks}{L} % Set of physical links.
\newcommand{\pcapacity}[1]{C(#1)} % Physical sensor capacity.
\newcommand{\pdomain}[1]{\mathcal{D}(#1)} % Physical sensor mapping domain.
\newcommand{\recdomain}[1]{D_{#1}} % Physical sensor mapping domain.
\newcommand{\plocation}[1]{\mathit{loc}(#1)} % Physical sensor node geographical location.
\newcommand{\trans}{T} % Transition matrix on a graph.
\newcommand{\radius}{r} % Wireless transmission radius of geometric random graph.
\newcommand{\degr}[1]{d_{#1}} % Degree of physical
\newcommand{\price}{B} % Sensor virtualization price matrix.
\newcommand{\hop}{h} % Hop count.
\newcommand{\maxH}{\bar{h}} % Maximum hop count.
\newcommand{\eps}{\epsilon} % Hop count.
\newcommand{\mapvar}{m}
\newcommand{\pricemin}[1]{b_{#1}^{\text{min}}}
\newcommand{\transpose}[1]{#1^\mathsf{T}}

\renewcommand{\acute}[1]{#1^{\prime}}

\begin{acronym}
\acrodef{sla}[SLA]{Service Level Agreement}
\acrodef{ls}[LS]{Least Squares}
\acrodef{admm}[ADMM]{Alternating Direction Method of Multipliers}
\acrodef{rade}[RADE]{Randomized and Asynchronous Distributed Estimation}
\acrodef{radv}[RADV]{Randomized and Asynchronous Distributed Virtualization}
\acrodef{mse}[MSE]{Mean Square Error}
\acrodef{ml}[ML]{maximum-likelihood}
\acrodef{iot}[IoT]{Internet of Things}
\acrodef{p2p}[P2P]{Peer-to-Peer}
\end{acronym}

\begin{document}

\title{Cloud-Assisted Remote Sensor Network Virtualization for Distributed Consensus Estimation}

\author{Sherif~Abdelwahab$^*$, Bechir Hamdaoui$^*$, and Mohsen Guizani$^{\dag}$
~\\
$^*$ \small Oregon State University, abdelwas,hamdaoui@eecs.oregonstate.edu~\\
$^{\dag}$ \small Qatar University, mguizani@ieee.org
}

\date{}
\maketitle

\begin{abstract}
We develop cloud-assisted remote sensing techniques for enabling distributed consensus estimation of unknown parameters in a given geographic area. We first propose a distributed sensor network virtualization algorithm that searches for, selects, and coordinates Internet-accessible sensors to perform a sensing task in a specific region.
The algorithm converges in linearithmic time for large-scale networks, and requires exchanging a number of messages that is at most linear in the number of sensors.
Second, we design an uncoordinated, distributed algorithm that relies on the selected sensors to estimate a set of parameters without requiring synchronization among the sensors.
Our simulation results show that the proposed algorithm, when compared to conventional ADMM (\acl{admm}), reduces communication overhead significantly without compromising the estimation error. In addition, the convergence time, though increases slightly, is still linear as in the case of conventional ADMM. 
\end{abstract}

\maketitle

\section{Introduction}
\label{sec:introduction}
As the \ac{iot} emerges, a rapid growth of the number of sensor-equipped things (e.g., smart-phones, tablets, etc.) is observed where it became important to rethink the way conventional sensor-based management techniques are designed.
Although sensor-based distributed sensing has already been investigated in the past, cloud-assisted remote sensing is a new paradigm that capitalizes on the capabilities of \ac{iot} to enable what is called \emph{Sensing as a Service} \cite{abdelwahab2014IoT}. Cloud-assisted remote sensing based distributed parameter estimation is one example of such services.

Traditional sensor networks depend primarily on the sophistication and accuracy of the sensory devices themselves to perform sensing tasks and meet quality of service requirements. In the Sensing as a Service model, cloud-based sensor networks rely mainly on swarms of participatory sensors to perform remote sensing tasks. Unlike traditional sensory devices, participatory sensors, though come with new opportunities, present key challenges, mainly pertaining to their sporadic availability and unpredictable mobility.

In the cloud-based remote sensing paradigm, a cloud agent (or manager) is responsible for receiving and handling remote sensing task requests from cloud clients. When a request is granted, the agent is also responsible for virtualizing a sensor network to perform the requested sensing task. An example of such tasks is distributed consensus estimation, in which an unknown set of parameters need to be estimated.
Imagine for example the virtual sensing task in which it is required to track the location of an RFID-tagged person in a large campus. A group of smart-phones connected through machine-to-machine physical links can estimate the location of a person based on the RFID signal strength each smart-phone receives and measures. In this case, a virtual sensing request is sent to the cloud agent which dispatches the request to few smart-phones in the swarm. These smart-phones autonomously search for a group of smart-phones that can read RFID tags and are willing to participate in the sensing task to estimate the distance based on received signal strengths.
Among all these smart-phones, a subgroup of them is then selected to form a virtual sensor network connected according to a topology that is to be specified by the cloud agent itself.% (see for example \figurename~\ref{fig:example}).
Such virtual sensor network distributedly and cooperatively estimates the location of the person using distributed linear consensus algorithms (see for example \cite{tarrio2011weighted,garin2010survey}), and continuously sends an update of the location of the person to the cloud agent, which in turn sends it to the cloud client.

In this paper, we develop cloud-assisted remote sensing algorithms that enable distributed consensus estimation of unknown parameters. Specifically, we propose:
\begin{itemize}
\item An efficient network virtualization algorithm that can search and select sensors from the swarm to form a virtual sensor network that can perform a sensing task. The algorithm consists of selecting a set of sensors that are willing to participate in the requested remote sensing task, and finding optimal one-to-one mappings between virtual and participatory sensors.

\item An efficient estimation algorithm that relies on the virtual sensor network, formed by our proposed virtualization algorithm, to estimate a set of unknown parameters in a distributed way and without requiring synchronization among participatory sensors.

\end{itemize}

Our results show that given a virtual sensing task requiring $\vnum$ sensors and a swarm of $\pnum$ sensors where a pair of sensors is considered connected if the sensors are within a distance $\radius$ from one another, our virtualization algorithm finds the capabilities of the sensors in $O(\radius^{-1}\log\pnum)$ with an average number of messages per sensor of $\Theta(1)$. Our results also show that our proposed algorithms achieve a virtualization benefit that is very close to an upper bound in
$ O( \max\{\radius^{-1}\pnum\log\pnum,\vnum^3\} ) $ time with a $O(\pnum)$ worst case average number of messages per sensor. Finally, our results show that our proposed distributed parameter estimation algorithm has a linear convergence time and incurs communication overhead that is at least an order of magnitude lesser than that incurred by the conventional \acf{admm}.

In our design, the cloud agent does not need to have full knowledge of the underlying swarm of sensors, including its topology, sensor capabilities, and sensor availability. We only assume that the cloud agent has a direct communication with some, but not necessarily all, sensors. Each sensor has to exchange small sized messages with only its direct neighbors to diffuse information across the whole or part of the swarm. We also assume that the swarm is dynamic by nature in that sensors' availability, locations, connectivity, and capacities may change sporadically and unpredictably over time.

This paper is organized as follows: In Section \ref{sec:model}, we overview our algorithmic design, describe the requirements, and state the related work. In Section \ref{sec:virtualization}, we propose and present our sensor network virtualization algorithm.
In Section \ref{sec:admm}, we propose and present our distributed parameter estimation algorithm.
Finally, we evaluate and compare the performance of our  proposed algorithms numerically in Section \ref{sec:results} and conclude the paper in Section \ref{sec:conclusion}.

\section{Framework Overview: System Model, Problem Formulation, and Related Work}
\label{sec:model}
In this section, we provide a brief overview of the different framework components and algorithms that are proposed in this paper to perform cloud-assisted remote sensing based parameter estimation. We also talk about some previously done works that are related to the algorithms proposed in this paper.

\subsection{Participatory Sensors}
We consider a swarm of a large number of participatory sensors that are managed by a cloud platform. Each sensor of the swarm is cloud accessible via the Internet and is assumed to have some sensing capability. We model the swarm as an Euclidean geometric random graph $\psn=(\psens,\plinks)$, where $\psens$ is a set of $\pnum$ sensors, and $\plinks$ is the set of all links connecting the sensors, where
two sensors are considered to be connected if they are within a transmission radius, $\radius$, of each other. Let $\plocation{i}$ denote the physical location of sensor $i$, and $\pcapacity{i}$ denote its sensing capability or capacity ($\pcapacity{i}$ can for e.g. refer to maximum allowed sensing time, maximum allowed processing power, maximum allowed memory capacity, etc.).

We assume that each sensor $i \in \psens$ is capable of estimating a vector of unknown parameters, $\param \in \reals^\paramsize$, through noisy measurements, $\obs_i \in \reals^\obssize$. That is,
$$\obs_i = \model_i \param + \noise_i,\; i=1,\ldots,\pnum$$
where $\model_i \in \reals^{\obssize \times \paramsize}$ is sensor $i$'s sensing model (typically      known to $i$ only) relating $\obs_i$ to $\param_i$, and $\noise_i$ is an additive Gaussian noise with zero mean and variance $\sigma_i^2$. We assume that $\noise_i$ and $\noise_j$ are independent from one another for all $i,j \in \psens$. Because different sensors may have different sensing models and/or different measurement methods, it is very likely that different sensors have different estimates of $\param$. Also, we do not assume/require that the sensors are synchronized; that is, the consensus algorithms we develop in this paper to estimate $\param$ are asynchronous.

\subsection{Virtual Remote Sensing}
We assume that there exists a cloud manager/agent that is responsible for managing the participatory sensors, handling virtual sensing task requests (to be submitted by cloud clients), and ensuring that clients' \acp{sla} are met once their requests are granted by the cloud.
Each virtual sensing task request is represented by a quadruple, $\langle \vnum,\tparam,\vsncenter,\vsndiam \rangle$, where $\vnum$ is the number of (virtual) sensors requested to perform the sensing task, $\tparam \in \reals^\paramsize$ is a column vector of unknown parameters to estimate by the virtual sensors, and
the location $\vsncenter$ and radius $\vsndiam$ indicate the area of interest that needs to be sensed; i.e., all requested sensors must be located within $\vsndiam$ distance from the center $\vsncenter$.
Generally speaking, \ac{sla}s consist of: (i) a maximum time within which the sensing task must be completed, (ii) an absolute tolerance $\eps_{\text{abs}} > 0$ of the estimation quality, (iii) a relative tolerance $\eps_{\text{rel}} > 0$ of the estimation quality (maximum gap between the $\vnum$ sensors' local estimates of $\tparam$), and (iv) a maximum rejection rate, defined as the ratio of the number of failed virtualizations to the total number of sensing task requests.

Upon receiving a sensing task request, the cloud agent's job is then to define a set $\vsens$ of $\vnum$ virtual sensors to be realized by $\vnum$ connected participatory sensors, all located within distance $\vsndiam$ from the center $\vsncenter$, that can collaboratively and distributively estimate $\tparam$. Depending on the \ac{sla}s of the sensing task, the cloud agent needs to determine each of the following three parameters before proceeding with sensor network virtualization needed to perform a requested sensing task. First, it needs to choose a suitable virtual topology that connects the set of virtual sensors, $\vsens$, so that they can perform the sensing task collaboratively. Although other topologies can be used, we focus in this paper on three types: complete, cyclic, and star. For a given topology, let $\vlinks$ denote the set of virtual links connecting the virtual sensors and $\vsn=(\vsens,\vlinks)$ be the graph representing the virtual sensor network. Note that the cloud agent needs to make sure that the virtual sensor network is connected according to the chosen topology.

Another parameter the cloud agent needs to fix and associate with $\vsn$ is the maximum allowed path length, $\maxH$, between any pair of virtual sensors. $\maxH$ can be viewed as a way to limit the number of sensors/hops message exchanges among virtual sensors can go through. It is a parameter that can be used to impose an upper bound on end-to-end message delays, and thus, on the sensing task completion time.
Note that a virtual link between two virtual sensors may be realized by more than two physical sensors, and some of these sensors may not necessarily realize a virtual sensor in $\vsens$ by itself. That is, some sensors may only be used for forwarding traffic without participating in performing the sensing task.

The third parameter the cloud agent needs to choose and set is the sensing capacity threshold $\vdemand{j}$ for a virtual sensor $j \in \vsens$. The capacity $\pcapacity{i}$ of a participatory sensor $i$ which realizes $j$ must then exceeds the capacity threshold $\vdemand{j}$. This threshold can for e.g. represents the minimum storage capacity, the minimum CPU computing power, and/or the minimum amount of time required by the sensing task.

The required values of the parameters associated with $\vsn$ including $\vlinks$, $\vdemand{j}$, and $\maxH$ will be mainly determined by the client's \ac{sla}, and is beyond the scope of this paper. In this paper, we focus instead on the design of efficient algorithms that meet these design requirements.
Our ultimate objective in this work is to design a distributed consensus algorithm that enables the estimation of the unknown parameter vector, $\param$, subject to the design parameter requirements specified by both the cloud client and the cloud agent. 
To this end, the key tasks that need to be executed by the swarm of participatory sensors to perform estimation are:

\subsubsection{Sensor Search}
It consists of searching for the sensors, among all participatory sensors, that meet the $\vsn$ requirements. More specifically, the swarm searches for a subset of participatory sensors, $\psens^\prime \subset \psens$, such that a sensor $i \in \psens^\prime$ if:
\begin{enumerate}[i)]
\item it can sense and estimate $\tparam$ through a sensing model $\model_i$; i.e., it can observe a vector $\obs_i$ that can be expressed as $\obs_i = \model_i \tparam + \noise_i$,

\item it is geographically located within $\vsndiam$ distance from $\vsncenter$, and

\item its capacity $\pcapacity{i} \geq \vdemand{j}$ for at least one virtual sensor $j \in \vsens$.
\end{enumerate}

For each participatory sensor $i$, we define the virtual domain of $i$, $\pdomain{i}$, as the set of all virtual sensors that can be supported by sensor $i$; that is,
\begin{equation}
\label{pdomain}
\pdomain{i} = \begin{dcases*}
        \{j \in \vsens : \pcapacity{i} \geq \vdemand{j}\} & if $\|\plocation{i} - \vsncenter \| \leq \vsndiam$\\
        \emptyset & otherwise,
                \end{dcases*}
\end{equation}
and the objective of this task is to construct, with the minimum possible communication overhead, each participatory sensor $i$'s virtual domain, $\pdomain{i}$, and to determine the set $\psens^\prime$ as fast as possible, all without assuming prior knowledge of the $\psn$ topology. Our proposed technique for performing such a task is presented in Section~\ref{sec:virtualization}.

\noindent{\bf Related work.}
In a recent work, Perera et. al \cite{perera2013context,perera2014sensor} described a system of context-aware sensor search to address the research challenges of searching for sensors when large numbers of sensors with overlapping and redundant functionality are available to the cloud. Such a sensor search approach suffers from practical limitations as it relies on centralized knowledge of all available sensors and requires continuous tracking of the sensors' dynamics, such as the sensors' availability, connectivity, and mobility.

Sensor search algorithms need to be simple, have bounded search latency, and incur minimum communication overhead between the cloud platform and the large number of participatory sensors.
Unlike centralized resource discovery algorithms \cite{belbekkouche2012resource}, gossip-based search protocols are distributed and topology-independent, which are more suitable for sensor search in the \ac{iot} context. Gossip protocols are originally designed for information dissemination \cite{birman2007promise}, and have been demonstrated to be effective in resource discovery in \ac{p2p} networks \cite{ferretti2013gossiping}. Our proposed framework relies on gossip techniques to determine the set of sensors among all participatory sensors that are capable of performing sensing and are willing to participate in the formation of a virtual sensor network.

\subsubsection{Sensor Network Virtualization}\label{subsubsec:virtualization}
This virtualization task consists of finding $(i)$ a set $\assign \subset \psens'$ of exactly $\vnum$ connected sensors selected among all sensors in $\psens'$ (the $\vnum$ selected sensors should be connected according to the virtual topology chosen by the cloud agent) and $(ii)$ a set $\mapf \subset \{(i,j) \in \assign \times \vsens: j \in \pdomain{i} \}$ of one-to-one mapped pairs (each participatory sensor in $\assign$ is mapped to one and only one virtual sensor in $\vsens$) such that the length, $\hop(i,i')$, of any simple path connecting two distinct participatory sensors $i,i'$ in $\assign$
mapping a pair of directly connected virtual sensors $(j,j') \in \vlinks$ is less than or equal to $\maxH$.
We refer to a possible $\{\assign,\mapf\}$ pair as a {\em feasible virtualization} of the requested virtual sensor network $\vsn$. Note that for any possible set $\assign$, there could exit multiple possible sets, $\mapf$, each can form a feasible virtulization when paired with $\assign$, and the objective of a sensor network virtualization algorithm is then to find the 'optimal' feasible virtualization, $\{\assign,\mapf\}^*$.

%\noindent{\bf Our proposed approach.}
We now define and introduce what an 'optimal' feasible virtualization means. We consider that the cost (to the cloud) of virtualizing a virtual sensor network request, $\vsn=(\vsens,\vlinks)$, is determined by the amounts of requested resources and given by
$ \text{Cost}(\vsn) =  \alpha |\vsens| + \beta |\vlinks|$, where $\alpha$ denotes an incentive paid by the cloud to each participatory sensor, and $\beta$ denotes an incentive associated with each physical path between each pair of participatory sensors in $\assign$. An incentive could be monetary or could be in any other form (e.g., credits, services, etc.).
On the other hand, the total benefit (to the cloud/swarm of sensors) resulting from a feasible virtulization, $\{\assign,\mapf\}$, can be expressed as
\begin{equation}
\label{totbenefit}
\text{Benefit} = \sum\limits_{(i,j) \in \mapf} \alpha \frac{\pcapacity{i}-\vdemand{j}}{\pcapacity{i}} 
 +  \sum\limits_{(i,i') \in \path} \beta \frac{\maxH-\hop(i,i')}{\maxH},
\end{equation}
where $\hop(i,i')$ is again the path length (in number of hops) of the path connecting the sensors of the pair $(i,i')$ mapping the virtual link between $j$ and $j'$ and $\path = \{(i,i') \in \assign \times \assign : (i,j),(i',j') \in \mapf\, , (j,j') \in \vlinks\}$ denotes the set of all such pairs.
Note that for a given sensing task request (i.e., for a given virtualization cost), the lesser the used physical resources, the higher the total benefit. The sensor network virtualization algorithm that we propose consists then of finding a feasible virtualization that maximizes the total benefit given in Eq.~\eqref{totbenefit}. We refer to the optimal solution as $\{\assign,\mapf\}^*$.
Clearly, finding $\{\assign,\mapf\}^*$ is a hard problem due to the factorial size of the solution space (in $\pnum$). Our first contribution in this work is to develop an algorithm that solves this virtualization problem efficiently.
%Our proposed algorithm is distributed and does not %require synchronization among sensors. 
The algorithm is presented in Section \ref{sec:virtualization}.

\noindent{\bf Related work.}
Network virtualization techniques proposed in the past decade consist mainly of virtual network embedding algorithms, which instantiate virtual networks on substrate infrastructures \cite{fischer2013virtual,belbekkouche2012resource}. Most of these virtual network embedding algorithms are centralized (e.g. \cite{cheng2011virtual}) due to the ease of deployment of centralized approaches in cloud platforms where the cloud provider desires to have full control on the physical network resources.

Distributed virtual network embedding and virtualization algorithms have also been proposed in literature~\cite{houidi2008distributed,beck2013distributed}. One of the limitation of the algorithm proposed in~\cite{houidi2008distributed} lies in its unsuitability for swarm virtualization, as the authors assume unlimited physical resources and consider an offline resource virtualization approach. As for the more recent work proposed in~\cite{beck2013distributed}, although the virtualization phase of the physical network does not require full/global knowledge of the swarm, the cloud must initially partition the swarm into hierarchies and delegate each virtual network request to a different hierarchy, which also requires full knowledge but about the sensors in each hierarchy.
Unlike this approach, our proposed virtualization algorithm is fully distributed. In addition, it does not require any synchronization among sensors. 

\subsubsection{Distributed Consensus Estimation} This task relies on the virtual sensor network (formed in the previous step) to provide an estimate of $\tparam$ that is at most $\eps_{\text{abs}}$ from the optimal, and that all the virtualized sensors consent to the same estimate value of $\tparam$ with a tolerance of $\eps_{\text{rel}}$.

Without loss of generality, consider indexing the selected $\vnum$ sensors in the virtual sensor network as $1\ldots\vnum$ and let
$\obs = \transpose{\left[\transpose{\obs_1},\ldots,\transpose{\obs_\vnum}\right]}$,
$\model = \transpose{\left[\transpose{\model_1},\ldots,\transpose{\model_\vnum}\right]}$, and
$\noise = \transpose{\left[\transpose{\noise_1},\ldots,\transpose{\noise_\vnum}\right]}$. The aggregate measurements can then be written as
$\obs = \model \tparam + \noise$.
One simple approach of estimating $\tparam$ is to have the cloud agent first collects from each virtual sensor $i$ its measurement vector, $\obs_i$, and its sensing model, $\model_i$, and then solves the following \ac{ls} problem
\begin{equation}
\begin{array}{ll}
\mbox{minimize}   & \frac{1}{2}\|\obs - \model \hat{\tparam}\|^2
\end{array}
\label{lsobjective}
\end{equation}
where $\hat{\tparam}$ is here the optimization variable. The unbiased \ac{ml} estimate of $\tparam$ is simply
$\hat{\tparam}_\text{LS} = \left(\transpose{\model} \model \right)^{-1} \transpose{\model} \obs$.

This centralized \ac{ls} approach, though simple, requires that each virtualized sensor exchanges its measurement vector and its sensing model with the cloud agent, which can create significant communication overhead, especially when the number of measurements, $\obssize$, and the number of virtual sensors, $\vnum$, are large.

We instead propose, in this paper, a decentralized approach that relies on the virtual sensor network to provide an estimation of the parameter vector $\tparam$. We rely on the recent results presented in \cite{Wei13onthe} to develop our distributed estimation algorithm, which reduces communication overhead significantly when compared to the conventional \ac{admm} approach~\cite{paul2013network} in addition to not requiring synchronization among sensors. 
The proposed algorithm is presented in Section~\ref{sec:admm}.

\noindent{\bf Related work.}
Distributed parameter estimation approaches have been proposed in~\cite{paul2013network,tarrio2011weighted,garin2010survey}. Estimation can for e.g. be carried out by first computing a local estimate at each virtual sensor and then perform a distributed weighted average of the local estimates~\cite{tarrio2011weighted}. This approach results in an \ac{ml} estimate, but does not limit/bound the variation between mean square errors of local estimates.
More recently, Paul et al. \cite{paul2013network} propose a distributed estimation algorithm based on \ac{admm}. Although this approach results in an optimal mean square error when compared to \ac{ls}, it exhibits a significant in-network communication overhead that requires even more messages to be exchanged among sensors than that exchanged in the centralized \ac{ls}. One approach also proposed in \cite{paul2013network} to overcome this problem is to approximate the computation of primal and dual variables at each step of the algorithm by using earlier versions of these variables instead of sharing them at each iteration which marginally reduces the communication overhead. In addition to the increased communication overhead, conventional \ac{admm} requires synchronous operation of the sensors. This is very challenging from a practical viewpoint, and does not scale well especially when applied in the \ac{iot} context. It has been shown recently that an asynchronous implementation of \ac{admm} has $O(1/\tindx)$ convergence \cite{Wei13onthe}.

Not only is our proposed estimation algorithm both asynchronous and distributed, but also reduces communication overhead significantly when compared to the conventional \ac{admm} approach~\cite{paul2013network}.

\section{Sensor Network Virtualization}
\label{sec:virtualization}
We begin by presenting our proposed \ac{radv} algorithm, which consists of four phases: $(I)$ searching for sensors that can support a virtual sensing task request $\vsn$, $(II)$ pruning of virtual domains $\pdomain{i}$ for all $i \in \psens$, $(III)$ construction of benefit matrices in a distributed manner, and $(IV)$ solving assignment problems at virtual sensors. This approach results in multiple solutions each evaluated by a different sensor, and the cloud agent selects the solution with the maximum benefit.

We design time-invariant gossip based algorithms for the first three phases in which sensors exchange information randomly~\cite{shah2009gossip,birman2007promise}. At the $\tindx\text{-th}$ time slot, let sensor $i$ be active and contact its neighbor sensor $j$ (i.e., $(i,j) \in \plinks$) with probability $\trans_{i,j} > 0$ only if $j$ can be contacted by more than one of its neighbors. The probability $\trans_{i,i}$ denotes the probability that $i$ does not contact any other sensor. Let the $\pnum \times \pnum$ matrix $\trans = [\trans_{i,j}]$ be a doubly stochastic transition matrix of non-negative entries \cite{boyd2006randomized}.
A natural choice of $\trans_{i,j}$ is
\begin{equation}
\label{transitionmat}
\trans_{i,j} = \begin{dcases*}
                             \frac{1}{\degr{i}+1}, & if $i = j$ or $(i,j) \in \plinks,$\\
                             0, & otherwise,
                           \end{dcases*}
\end{equation}
where $\degr{i} = |\{j \in \psens : (i,j) \in \plinks \}|$ is the degree of sensor $i$.

We now present each of the four phases in greater details.

\subsubsection*{\textbf{Phase I}---Sensor Search}
The objective of this phase is to construct $\pdomain{\pnode}$ for all $\pnode \in \psens$ and $\psens^\prime$ as fast as possible without assuming prior knowledge of the $\psn$ topology.
First, the cloud initiates the search at time $\tindx=0$ by sending $\vsn$ to one or more arbitrary sensors. At any later time $\tindx$, $\pnode$ and its neighbor $\acute{\pnode}$ exchange information as follows. $\pnode$ pushes $\vsn$ to $\acute{\pnode}$ only if $\acute{\pnode}$ does not have $\vsn$, or pulls it from $\acute{\pnode}$ only if $\pnode$ does not have $\vsn$.
%such that $\preceived(\tindx) = \preceived(\tindx-1) \cup \{\acute{\pnode}\}$.
%such that $\preceived(\tindx) = \preceived(\tindx-1) \cup \{i\}$.
If $\pnode$ contacts $\acute{\pnode}$ and both $\pnode$ and $\acute{\pnode}$ have received $\vsn$ before, $\pnode$ stops contacting any other sensor. Upon receipt of $\vsn$, a  sensor $\pnode$ constructs $\pdomain{\pnode}$ according to Eq.\eqref{pdomain} and $\psens^\prime$ as
$$\psens^\prime = \{\pnode \in \psens : |\pdomain{\pnode}| > 0\}.$$

A virtual domain of sensor $i$, $\pdomain{i}$, evaluated during the sensor search phase is not sufficient to tell whether a feasible embedding can be found if sensor $i$ virtualizes (is mapped to) $j$. For example, the virtualization in which sensor $i$ virtualizes $j$ and sensor $\acute{i}$ virtualizes $\acute{j}$ where there is a virtual link $(j,\acute{j})$ is not feasible if $\acute{i}$ is not reachable from $i$ and vice versa. When this situation occurs, we say that sensor $i$ is incapable of supporting the topology requirement specified by $\vlinks$, initiating thus a virtual domain pruning.

\subsubsection*{\textbf{Phase II}---Virtual Domain Pruning}
During this phase, we ensure that all virtualized sensors maintain the topology $\vlinks$ by allowing a sensor to receive the virtual domains of other sensors and delete a virtual sensor $j$ from its domain if there exists a virtual link $(j,\acute{j})$ such that $\acute{j}$ is not included in any other received domains.
Let
$\recdomain{\pnode} \subset \{\pdomain{i} : i\in\psens \}$
%$\recdomain{\pnode} \subset %\bigcup\limits_{i\in\psens} \pdomain{i}$
denotes the set of domains that sensor $\pnode$ has at time $\tindx$. Initially $\recdomain{\pnode} = \{\pdomain{\pnode}\}$ and $\hop(i,\pnode) = 0\, \text{for all}\, i \in \psens$\footnote{Knowledge about other sensors existence is not needed, and $\hop$ is typically evaluated dynamically.}.
Using the same transition matrix, $\trans$, defined in Eq.~\eqref{transitionmat}, $\pnode$ contacts only one of its neighbors $\acute{\pnode}$ at time $\tindx$.
Then, $\text{for all} \,\pdomain{i} \in \recdomain{\pnode}\,:\,i \neq \acute{\pnode}$, $\pnode$ pushes $\pdomain{i}$ to $\acute{\pnode}$ only if $\acute{\pnode}$ did not receive $\pdomain{i}$ before and $\hop(i,\pnode) < \maxH.$
Also, $\text{for all} \,\pdomain{i} \in \recdomain{\acute{\pnode}}\,:\,i \neq \pnode$, $\pnode$ pulls $\pdomain{i}$ from $\acute{\pnode}$ only if $\pnode$ did not receive $\pdomain{i}$ before and $\hop(i,\acute{\pnode}) < \maxH.$
If no information is exchanged between $\pnode$ and $\acute{\pnode}$ at time $\tindx$, $\pnode$ stops contacting any of its neighbors. However, $\pnode$ may restart contacting its neighbors again if it updated $\recdomain{\pnode}$ after time $\tindx+1$.

When $\pnode$ constructs its $\recdomain{\pnode}$, it starts by pruning $\pdomain{\pnode}$. The pruning is performed by deleting a virtual sensor $j \in  \pdomain{\pnode}$ (i.e.,
$\pdomain{\pnode} \leftarrow \pdomain{\pnode} \setminus \{j\}$) if none of the virtual sensors that are connected to $j$, $\{\acute{j} \in \vsens: (j,\acute{j}) \in \vlinks\}$, is not included in any received $\pdomain{i}$,
i.e. $j \notin \pdomain{i}: \pdomain{i} \in \recdomain{\pnode}.$ This pruning rule ensures that the virtualized sensors maintain the required topology $\vlinks$ and the constructed benefit matrices shall result in a feasible virtualization.

\subsubsection*{\textbf{Phase III}---Construction of Benefit Matrices}
As mentioned earlier, finding a feasible virtualization, $\{\assign,\mapf\}^*$, that maximizes the total benefit given in Eq.~\eqref{totbenefit} is a hard problem due to the large size of the solution space. Therefore, this phase proposes an efficient way of solving this virtualization problem. Specifically, we propose a method that solves this problem in a distributed manner and without requiring any synchronization among sensors, as described next.

During this phase, each sensor $\pnode$ locally constructs its own set, $\assign^{(\pnode)}$, of $\vnum$ sensors that $\pnode$ chooses as virtualized sensors to assign to virtual sensors in $\vsens$.
Each sensor $\pnode$ also maintains $\vnum$ row vectors, $\price_{i}^{(\pnode)} \in \reals^{1 \times \vnum}\,\text{and } i\in \assign^{(\pnode)}$, that we define as the benefit vector of sensor $i$ seen by $\pnode$, where the $j\text{-th}$ element, $\price_{i,j}^{(\pnode)}$, denotes the benefit of assigning participatory sensor $i \in \assign^{(\pnode)}$ to the virtual sensor $j \in \vsens$ as seen by $\pnode$, and is given by
\[
\price_{i,j}^{(\pnode)} = \begin{dcases*}
                             \alpha \frac{\pcapacity{i} - \vdemand{j}}{\pcapacity{i}} + \beta \frac{\maxH - \hop(j,\pnode)}{\maxH} & if $j \in \pdomain{i},$\\
                             0 & otherwise.
                           \end{dcases*}
\]

Our objective is then to construct, for each $\pnode \in \psens$, the benefit matrix
$\price^{(\pnode)} = [\price_{i \in \assign^{(\pnode)}}^{(\pnode)}]$
as fast as possible, and find a feasible virtualization, $\{\assign,\mapf\}$, that maximizes the total benefit,
$$\sum\limits_{(i,j) \in \mapf} \price_{i,j}^{(\pnode)},$$
among all $\pnode \in \psens$ without knowing the $\psn$ structure. Moreover, the path length between a sensor $\pnode$ and any other sensor $i$ that $\pnode$ includes in its benefit matrix must not exceed $\maxH$. Finally, a sensor $\pnode$ shall include only the benefit vectors of the $\vnum$ sensors with the largest possible benefit.

Each sensor $\pnode$ initially sets $\assign^{(\pnode)} = \assign^{(\pnode)} \cup \{\pnode\}$ if $\pdomain{\pnode} \notin \emptyset$, sets $\hop(i,\pnode) = 0 \, \text{for all}\, i \in \psens$, and sets
\[
\price_{\pnode,j}^{(\pnode)} = \begin{dcases*}
                                   \alpha \frac{\pcapacity{s} - \vdemand{j}}{\pcapacity{s}} + \beta, & $j \in \pdomain{\pnode},$\\
                                    0, & otherwise.
                           \end{dcases*}
\]
Also, $\pnode$ maintains a scalar, $\pricemin{\pnode}$, defined as the minimum total benefit it has received from any other  sensor and written as
$$\pricemin{\pnode} = \min_{i} \sum\limits_{j \in \vsens} \price_{i,j}^{(\pnode)} ,$$ and the corresponding sensor,
$$i_{\pnode}^\text{min} = \argmin_{i} \sum\limits_{j \in \vsens} \price_{i,j}^{(\pnode)}.$$
Initially, $\pricemin{\pnode}=0$ and remains so until $|\assign^{(\pnode)}| = \vnum$.

Using the same transition matrix, $\trans$, defined in Eq.~\eqref{transitionmat}, $\pnode$ contacts its neighbor $\acute{\pnode}$ only once at each time $\tindx$.
Then, $\text{for all} \,i \in \assign^{(\pnode)}\,:\,i \neq \acute{\pnode}$, $\pnode$ pushes the benefit vector $\price_{i}^{(\pnode)}$ to $\acute{\pnode}$ only if $\hop(i,\pnode) < \maxH$ and
$$\sum\limits_{j \in \vsens} \left( \price_{i,j}^{(\pnode)}-\frac{\beta}{\maxH} \right) > \pricemin{\acute{\pnode}}.$$
Also, $\text{for all} \,i \in \assign^{(\acute{\pnode})}\,:\,i \neq \pnode$, $\pnode$ pulls the benefit vector $\price_{i}^{(\acute{\pnode})}$ from $\acute{\pnode}$ only if $\hop(i,\acute{\pnode}) < \maxH$ and
$$\sum\limits_{j \in \vsens} \left( \price_{i,j}^{(\acute{\pnode})}-\frac{\beta}{\maxH} \right) > \pricemin{\pnode}.$$
If no information is exchanged between $\pnode$ and $\acute{\pnode}$ at time $\tindx$, $\pnode$ stops contacting its neighbors at time $\tindx+1$. However, $\pnode$ may restart contacting its neighbors again if $\price^{(\pnode)}$ is updated after time $\tindx+1$.

When $\pnode$ receives $\price_{i}^{(\acute{\pnode})}$, $\pnode$ updates $\price_{i,j}^{(\pnode)}$ as
\[
\price_{i,j}^{(\pnode)} = \begin{dcases*}
                                   \price_{i,j}^{(\pnode)} - \frac{\beta}{\maxH} & \mbox{ if } $j \in \pdomain{i},$\\
                                    0 & otherwise.
                            \end{dcases*}
\]

If $i \notin \assign^{(\pnode)}$, then we have two scenarios.
In the first scenario, $\pnode$ still has not received $\vnum$ benefit vectors, so $\pricemin{\pnode}=0$ and $|\assign^{(\pnode)}| < \vnum$, then $\pnode$ updates its set of candidate sensors as $\assign^{(\pnode)} = \assign^{(\pnode)} \cup \{i\}.$
In the other scenario in which $|\assign^{(\pnode)}| = \vnum$, $\pnode$ replaces the  sensor corresponding to the minimum total benefit, $i_{\pnode}^\text{min}$, with $i$ so that
$\assign^{(\pnode)} = \assign^{(\pnode)} \setminus \{i_{\pnode}^\text{min}\} \cup \{i\}.$
On the other hand, if $i \in \assign^{(\pnode)}$, then $\pnode$ updates $\price_{i,j}^{(\pnode)}$ if
$\sum\limits_{j \in \vsens} \price_{i,j}^{(\acute{\pnode})} > \sum\limits_{j \in \vsens} \price_{i,j}^{(\pnode)}.$
Finally, $\pnode$ updates $\pricemin{\pnode}$, $i_{\pnode}^\text{min}$, and
$\hop(i,\pnode)$ as
$\hop(i,\pnode)=\hop(i,\acute{\pnode})+1.$

Finding a feasible virtualization that maximizes the benefit
$\price^{(\pnode)} = [ \begin{array}{ccc}
\price_{i \in \assign^{(\pnode)}}^{(\pnode)}
\end{array} ]$
instead of the benefit given in Eq.~\eqref{totbenefit} makes the problem easier because every sensor has a different value for the benefit $\price_{i,j}$ that depends only on the length of the physical path between $i$ and $\pnode$ instead of the path lengths of all possible combinations of sensor pairs $(i,\acute{i})$ that can virtualize a virtual link. Intuitively, this relaxation still leads to an optimal or near optimal virtualization, because for a connected swarm $\psn$, the number of sensors that are directly connected by a single physical link (clique) grows logarithmically in $\pnum$ and hence this number is larger than $\vnum$ almost surely as $\vnum \ll \pnum$. In such a case, it is sufficient to ensure that the length of the paths between $i$ and $\pnode$ and between $\acute{i}$ and $\pnode$ are the shortest possible ones to ensure that the length of the path between $i$ and $\acute{i}$ is also the shortest, as in this case, $\pnode$, $i$, and $\acute{i}$ reside in the same clique with high probability. We evaluate the effectiveness of this relaxation in Section \ref{sec:results} and show that our virtualization algorithm performs well even when the condition $\vnum \ll \pnum$ does not hold.

\subsubsection*{\textbf{Phase IV}---Solving Local Assignment Problem}
After reception of the $\vnum$ benefit vectors, $\pnode$ proceeds to this phase of the algorithm only if it stops communicating and $|\assign^{(\pnode)}| = \vnum$.	
%\subsubsection{Solving an assignment problem}
Each  sensor $\pnode \in \psens$ with $|\assign^{(\pnode)}| = \vnum$ solves locally the following assignment problem:
\begin{equation}
\begin{array}{ll}
\mbox{maximize}   & \sum\limits_{i \in \assign^{(\pnode)}}\,\sum\limits_{j \in \pdomain{i}} \price_{i,j}^{(\pnode)} \mapvar_{ij}   \\
\mbox{subject to} & \sum\limits_{j \in \pdomain{i}} \mapvar_{ij} = 1, \; i \in \assign^{(\pnode)},  \\
				  & \sum\limits_{\{i: j \in \pdomain{i}\}} \mapvar_{ij} = 1, \; j \in \vsens,  \\
  				  & \mapvar_{ij} \in \{0,1\}, \\
\end{array}
\label{assignprob}
\end{equation}
where $\mapvar_{ij}$ are binary optimization variables indicating whether the participatory sensor $i$ is assigned to the virtual sensor $j$.
The problem formulated in \eqref{assignprob} is equivalent to the maximum weight matching perfect problem in a bipartite graph, and hence, we propose to use the classical Hungarian method to solve it (the worst case time complexity is $O(\vnum^3)$~\cite{kuhn1955hungarian,munkres1957algorithms}).

We can also tolerate an error $\epsilon > 0$ of the resulting total benefit and relax the restriction of finding a perfect matching for large $\vnum$. This relaxation is reasonable when there are enough sensors involved in solving these local optimization problems, as in this case we can pick the best solution and discard those without a perfect matching. In such a scenario, we can also use a linear time $(1-\epsilon)\text{-approximation}$ algorithm to solve \eqref{assignprob} \cite{duan2014linear}. In this paper, we use the Hungarian method to solve our formulated optimization problems. Details of the algorithm are omitted due to space limitation; readers are referred to~\cite{kuhn1955hungarian,munkres1957algorithms,duan2014linear} for detailed information.

Each sensor solves locally the optimization problem given in~\eqref{assignprob} and sends its obtained solution to the cloud agent. This is done asynchronously. The cloud agent then selects the solution that leads to the maximum total benefit, and keeps all other solutions for later use in the event that the network dynamics invalidate the selected solution before the virtual sensing task completes.

\noindent{\bf Complexity and message overhead.}
The time required to spread $\vsn$ across the network is $O(\radius^{-1}\log\pnum)$~\cite{shah2009gossip}. It takes $O(\vnum)$ worst case time to evaluate $\pdomain{i}$ locally at sensor $i$. Also, the time required to spread information in the pruning and benefit construction phases is $O(\radius^{-1}\pnum\log\pnum)$. The pruning of the virtual domain $\pdomain{i}$ requires node $i$ to examine $\vnum$ received virtual domains, each having at most $\vnum$ entries. The  worst case local running time of pruning is then $O(\vnum^2)$. Finally, the local running time of the Hungarian method is $O(\vnum^3)$. Hence, the overall complexity of is  $O(\max\{\radius^{-1}\pnum\log\pnum,\vnum^3\})$.

\begin{figure}[htp]
\centering
\includegraphics[width=0.38\textwidth]{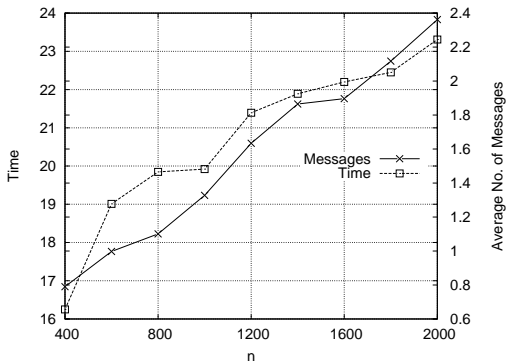}
\caption{Time (in number of iterations) and message overhead (in number of communicated messages) resulting from constructing the benefit matrices.}
\label{fig:radv_opt}
\end{figure}

The average number of messages communicated per sensor during the sensor search phase is $\Theta(1)$ and each message is $O(\vnum)$ in size. During pruning of virtual domains, since every sensor exchanges a maximum of $\pnum$ domains each of size that is also $O(\vnum)$, the average number of messages communicated per sensor is $O(\pnum)$. However, because we restrict that messages be communicated up to $\maxH$ hops for only a group of sensors that support the requirements of $\vsn$, the average number of messages per sensor is typically small. \figurename~\ref{fig:radv_opt} shows the total time and the average number of messages per sensor required during both the domain pruning and the benefit construction phases.
The total time growth is linearithmic in $\pnum$ when $\vsn$ is sent to exactly one sensor and when $\psn$
is connected. This time can, in practice, be decreased significantly if $\vsn$ is initially sent to multiple sensors.
%The cloud agent can simply choose these sensors %randomly or based on any sensor partition algorithm.
Additionally, the average number of messages per sensor is shown to scale linearly with $\pnum$, and is typically a very small fraction of $\pnum$.

\section{Distributed Estimation of $\tparam$}
\label{sec:admm}
After completing the sensor virtualization task, using the proposed \ac{radv} algorithm that we described in Section \ref{sec:virtualization}, the virtual sensors run an in-network parameter estimation algorithm to compute $\hat{\tparam}$ distributedly. In this section, we present our proposed \ac{rade} algorithm.
We first follow the standard \ac{admm} approach to derive primal, dual and Lagrangian variable update equations, then we describe the proposed \ac{rade} algorithm. For clarity of notation, in what follows, we refer to the set of $\vnum$ selected participatory sensors, determined by means of the proposed \ac{radv} algorithm, simply as $\assign$. 

The centralized estimation approach given in \eqref{lsobjective} is first decomposed into $\vnum$ local estimates of $\tparam$ (one $\hat{\tparam}_i\; \mbox{for each } i \in \assign$) while constraining the local estimates with the coupling constraints ${\tparam}_i = {\tparam}_j\,\text{for all}\,(i,j) \in \path$. This results in the following optimization problem:
\begin{equation}
\begin{array}{ll}
\mbox{minimize}   & \frac{1}{2}\sum\limits_{i \in \assign} \|\obs_i - \model_i {\tparam}_i\|^2  \\
\mbox{subject to} & {\tparam}_i-{\tparam}_j=0\text{ for all }  (i,j) \in \path,
\end{array}
\label{consensusobj}
\end{equation}
where $\{{\tparam}_i, i \in \assign\}$ are the optimization variables.

By introducing an auxiliary variable, $\aux$, we decouple the constraints in \eqref{consensusobj}, so that ${\tparam}_i-\aux=0\;\text{for all} \;  i \in \assign$ \cite{palomar2006tutorial}. However, this requires that $\aux$ be shared among all $\vnum$ sensors. Instead, we introduce $\vnum$ auxiliary variables, $\aux_i$, and equivalently write the optimization problem as
\begin{equation}
\begin{array}{ll}
\mbox{minimize}   & \frac{1}{2}\sum\limits_{i \in \assign} \|\obs_i - \model_i {\tparam}_i\|^2  \\
\mbox{subject to} & {\tparam}_j-{\aux}_i=0\;\text{for all} \;  (i,j) \in \path.
\end{array}
\label{distobj}
\end{equation}
Let $\lagrmul = \{\lagrmul_{i,j} \in \reals^{\paramsize \times 1}: (i,j) \in \path\}$
and $\lagrpen = \{\lagrpen_{i,j} \in \reals : (i,j) \in \path\}$ denote respectively the set of Lagrangian multipliers and the set of penalty parameters. The augmented Lagrangian is
\begin{equation}
\label{auglagr}
\begin{array}{ll}
\lagr_{\lagrpen}(\tparam,\aux,\lagrmul) =  \sum\limits_{i \in \assign}& \Bigg[  \frac{1}{2} \|\obs_i - \model_i {\tparam}_i\|^2\\
						 & - \sum\limits_{j \in \assign : (i,j) \in \path} \transpose{\lagrmul_{i,j}} ({\tparam}_i - \aux_j)          \\
						 & + \sum\limits_{j \in \assign : (i,j) \in \path} \frac{\lagrpen_{i,j}}{2} \|{\tparam}_i - \aux_j\|^2 \Bigg].
\end{array}
\end{equation}
By setting the gradient w.r.t ${\tparam}_i$ of Eq.~\eqref{auglagr} to zero and solving for ${\tparam}_i$, we get
$$
\begin{array}{ll}
{\tparam}_i & = \left(\transpose{\model_i} \model_i + \sum\limits_{j \in \assign : (i,j) \in \path} \lagrpen_{i,j} \eyem  \right) ^ {-1}\\
	& .  \left( \transpose{\model_i} {\obs}_i +  \sum\limits_{j \in \assign : (i,j) \in \path} \left( \lagrmul_{i,j} + \lagrpen_{i,j} \aux_j \right)\right).
\end{array}
$$
Similarly, we solve for $\aux_i$ by setting the gradient w.r.t to $\aux_i$ to zero and rearranging the indices of the Lagrangian multipliers and the penalty parameters. It follows that
$$
\aux_i = \frac{1}{\vnum} \sum\limits_{j \in \assign : (i,j) \in \path} \left({\tparam}_j - \frac{1}{\lagrpen_{j,i}} \lagrmul_{j,i} \right).
$$

The former analysis leads to the conventional \ac{admm}-based distributed consensus estimation algorithm given by
\begin{equation}
\label{eq:admm}
\begin{array}{ll}
{\tparam}_i^{(\tindx+1)} & = \left(\transpose{\model_i} \model_i + \sum\limits_{j \in \assign : (i,j) \in \path} \lagrpen_{i,j} \eyem  \right) ^ {-1}\\
	& .  \left( \transpose{\model_i} {\obs}_i +  \sum\limits_{j \in \assign : (i,j) \in \path} \left( \lagrmul_{i,j}^{(\tindx)} + \lagrpen_{i,j} \aux_j^{(\tindx)} \right)\right),\\

\aux_i^{(\tindx+1)} & = \frac{1}{\vnum} \sum\limits_{j \in \assign : (i,j) \in \path} \left({\tparam}_j^{(\tindx)} - \frac{1}{\lagrpen_{j,i}} \lagrmul_{j,i}^{(\tindx)} \right),\\

\lagrmul_{j,i}^{(\tindx+1)} & = \lagrmul_{j,i}^{(\tindx)} - \lagrpen_{j,i} \left( {\tparam}_j^{(\tindx+1)} - \aux_i^{(\tindx+1)} \right),

\end{array}
\end{equation}
where the superscript $\tindx$ denotes the value of the variable at the $\tindx\text{-th}$ iteration. This conventional \ac{admm} algorithm, given in \eqref{eq:admm}, requires synchronization and variable update among the sensors~\cite{BoD:11,xu2014fast}. Moreover, at each iteration $\tindx$, each sensor $i$ must send $\aux_i^{(\tindx)}$ and ${\tparam}_i^{(\tindx)}$ to all other sensors it is connected to, so as to evaluate their $\tindx+1$ primal, dual, and Lagrangian multipliers. When $\obssize$ is small, this algorithm incurs communication overhead that can be shown to be worse than the communication overhead incurred by centralized estimation methods. However, when $\obssize$ is large, the conventional \ac{admm} algorithm incurs lesser communication overhead than what centralized estimation methods incur, but it still remains practically unattractive due to other weaknesses, detailed later in Section \ref{sec:results}.

Given the absolute and relative tolerances, $\eps_{\text{abs}}$ and $\eps_{\text{rel}}$, specified by the \acp{sla} between the cloud client and the cloud agent, we define the primal and dual tolerances, controlling the convergence of the algorithm at iteration $\tindx$, as
$$
\eps_i^{pri}(\tindx) = \sqrt{\vnum} \eps_{\text{abs}} + \eps_{\text{rel}} \max (\|{\tparam}_i^{(\tindx)}\|,\|-\aux_i^{(\tindx)}\|),
$$
and
$$
\eps_i^{dual}(\tindx) = \sqrt{\vnum} \eps_{\text{abs}} + \eps_{\text{rel}} \sum\limits_{j \in \assign} \|\lagrpen_{j,i} \lagrmul_{j,i}\|.
$$
The tolerances, $\eps_i^{pri}$ and $\eps_i^{dual}$, define the stopping criteria of sensor $i$; i.e., sensor $i$ stops updating  ${\tparam}_i$ and $\aux_i$ when
\begin{equation}
\label{pristop}
\| {\tparam}_i^{(\tindx+1)} - \aux_i^{(\tindx+1)} \| < \eps_i^{pri}(\tindx),
\end{equation}
and
\begin{equation}
\label{dualstop}
\| \aux_i^{(\tindx+1)} - \aux_i^{(\tindx)} \| < \eps_i^{dual}(\tindx).
\end{equation}
The stopping criteria of \ac{rade} are different from those of the conventional \ac{admm}. Unlike the conventional \ac{admm} where all sensors shall stop computations all at the same time using a common stopping criteria and common primal and dual tolerances, the stopping criteria \eqref{pristop} and \eqref{dualstop} of \ac{rade} allow a sensor $i$ to stop its computations asynchronously and independently from other sensors. However, these criteria are not enough to ensure asynchronous implementation, as synchronization is still required for dual and primal variable updates at iteration $\tindx+1$ due to their dependencies on $\tindx$.

To ensure full asynchronous implementation, we use the doubly stochastic transition matrix, $\trans \in \reals^{\vnum \times \vnum}$, where $\trans_{i,j}$ is the probability that a sensor $i$ contacts another  sensor $j$ at any iteration, for deciding the communications among sensors. We can have
\begin{equation*}
\trans_{i,j} = \begin{dcases*}
                             \frac{1}{\degr{i}'+1} & if $i=j$ or $(i,j) \in \path,$\\
                             0 & otherwise,
                           \end{dcases*}
\end{equation*}
where $\degr{i}' = |\{j \in \assign : (i,j) \in \path \}|$ is the degree of the virtual sensor, in $\vsn$, that $i$ virtualizes.
At iteration $\tindx+1$, sensor $i$ may need to contact only one sensor $j$, unless both of $i$'s stopping criteria, \eqref{pristop} and \eqref{dualstop}, are already satisfied. Whereas sensor $j$ can be contacted by more than one sensor if $j$ is not contacting any other sensor, even when both of $j$'s stopping criteria are satisfied.

Upon contacting $j$, sensor $i$ pushes ${\tparam}_i^{(\tindx)}$ to $j$ only if $i$'s primal stopping condition is not satisfied and pushes $\aux_i^{(\tindx)}$ to $j$ only if $i$'s dual stopping condition is not satisfied. Also, $i$ pulls ${\tparam}_j^{(\tindx)}$ from $j$ only if $j$'s primal stopping condition is not satisfied and pulls $\aux_j^{(\tindx)}$ only if $j$'s dual stopping condition is not satisfied. Finally, both $i$ and $j$ update their $\tindx+1$ variables using the most recent values they received from other sensors.

\begin{figure}%[htp]
\centering
\includegraphics[width=0.38\textwidth]{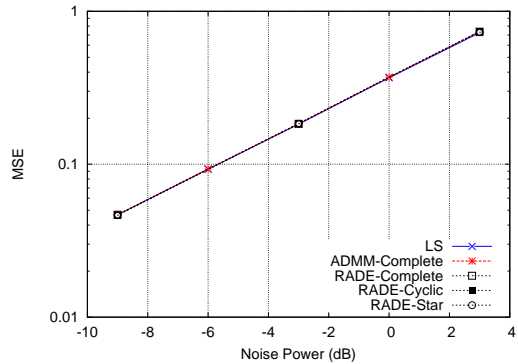}
\caption{\ac{mse} of \ac{rade} compared to those achieved under \ac{admm} and \ac{ls} at different noise power and for different virtual sensor network topologies.}
\label{fig:error_topology}
\end{figure}

\noindent{\bf Mean square error and convergence.}
The asynchronous and randomization design features of \ac{rade} do not impact the \ac{mse} achieved by \ac{rade} when compared to \ac{admm}. This is explained as follows. In both \ac{admm} and \ac{rade}, the number of necessary dual and primal variables updates required until convergence remains unchanged, so that convergence to the same estimate is guaranteed in both algorithms. \figurename~\ref{fig:error_topology} shows the \ac{mse} achievable under both \ac{rade} and \ac{admm} when compared to \ac{ls} under each of the three studied sensor network topologies: complete, star, and cycle. These results show the optimality of \ac{rade} that we intuitively discussed. All approaches have the same accuracy. But of course each of them does so at a different performance cost, as will be discussed later.

\begin{figure}[htp]
\centering
\includegraphics[width=0.38\textwidth]{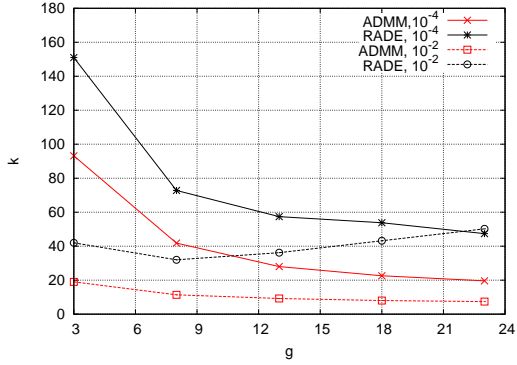}
\caption{Number of time steps (k) needed until convergence of \ac{rade} when compared to \ac{admm} for complete topology under different relative tolerance values $\epsilon_{\text{rel}}$.}
\label{fig:Convergence_VN}
\end{figure}

% asynchronous admm
On the other hand, \ac{rade} exhibits a linear convergence rate ($O(1/\tindx)$), similar to what the conventional \ac{admm} does. \figurename~\ref{fig:Convergence_VN} shows the number of time steps required for both \ac{rade} and \ac{admm} to converge under different relative tolerance parameters, $\eps_{\text{rel}}$. \ac{rade} convergence tends to be more restricted by the randomization nature of the algorithm for smaller values of $\eps_{\text{rel}}$, which can be seen by the increasing number of steps as $\vnum$ increases if $\eps_{\text{rel}} = 10^{-2}$. \ac{admm} generally requires a lesser number of steps to converge by relaxing the consensus constraint (through reducing $\eps_{\text{rel}}$). However, as will be seen in the numerical results section later, this increase in the number of convergence steps is acceptable when considering the amount of communication overhead that the algorithm saves.

\section{Numerical Results}
\label{sec:results}
In this section, we evaluate the performance of the proposed \ac{radv} and \ac{rade} algorithms through simulations. In our simulations, the swarm of sensors, $\psn$, and the virtual sensing task requests, $\vsn$, are generated using the parameters summarized in Table~\ref{sim_param}. The virtual sensor network topology can either be complete, cyclic, or star, with a randomly chosen central location, $\vsncenter$. We consider receiving and servicing only one virtual sensing task request at a time. The absolute and relative tolerances, $\eps_{\text{abs}}$ and $\eps_{\text{rel}}$, are set to $10^{-4}$ unless specified otherwise.	

\begin{table}[htp!]
%\vspace{15pt}
\renewcommand{\arraystretch}{1.3}
\caption{Simulation Parameters}
\label{sim_param}
\centering
\begin{tabular}{|l||l|l|l|l|l|}
\hline
\bfseries Parameter & $\radius$ & $\pcapacity{i}$ & $\vdemand{j}$ & $\vsndiam$ & $\maxH$  \\
\hline\hline
\bfseries Value & $0.1$ & $\sim U(50,100)$ & $\sim U(25,50)$ & $0.2$ & $20$ \\
\hline
\end{tabular}
%\vspace{-10pt}
\end{table}

\begin{figure}%[htp]
\centering
\includegraphics[width=0.38\textwidth]{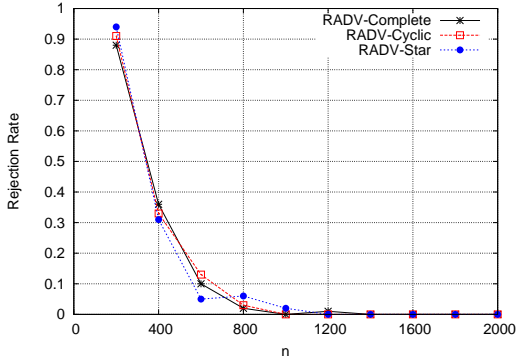}
\caption{Rejection rate encountered at different $\pnum$.}
\label{fig:radv_rejection}
\end{figure}

\figurename~\ref{fig:radv_rejection} shows the rejection rate encountered with different $\vsn$ topologies and $\pnum$ values. As we only consider one single request at a time, the results shown in this figure reflect mainly the impact of the virtual sensor network topology, the number of sensors $\pnum$, and the simulations parameters given in Table \ref{sim_param} on the rejection rate. The denser the swarm of sensors is, the lower the rejection rate is, implying that the cloud is capable of granting higher number of requests. As a marginal note, a star topology is slightly easier to virtualize than a complete or a cyclic topology.

One way of assessing the effectiveness of the virtualization algorithm is by measuring the difference between the total virtualization benefit given in \eqref{totbenefit} and the cost associated with the sensor virtualization introduced in Section~\ref{subsubsec:virtualization}. For a given number of virtual sensors, the cost is mainly determined by the choice of the topology (star topology has the lowest cost and complete topology has the highest one).
For a given topology, the total benefit is maximized when each virtual sensor is assigned to the participatory sensor with the maximum capacity and each virtual link is mapped to exactly one physical link. We refer to this maximized benefit as the upper bound.

\begin{figure}%[htp]
\centering
\includegraphics[width=0.38\textwidth]{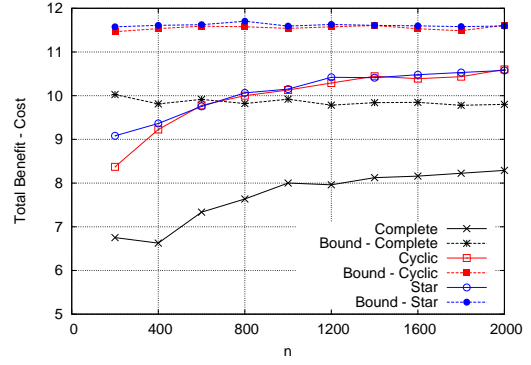}
\caption{Virtualization cost of \ac{radv} when compared to the upper bound under different topologies.}
\label{fig:radv_optval}
\end{figure}

In \figurename~\ref{fig:radv_optval}, we evaluate the virtualization effectiveness achieved by \ac{radv} under different virtual topologies. As the swarm gets denser, \ac{radv} achieves a  $\text{Total Benefit} - \text{Cost} $ that is very close to the upper bound. Since the lowest possible virtualization cost is with star or cyclic topologies, it is desired by the cloud to always arrange each virtual sensing task in a star or a cyclic topology. On the other hand, convergence and communication overhead of the distributed estimation is also impacted by the cloud agent's choice of the virtual topology. This creates a design trade-off, as we will see in the next two paragraphs.

\begin{figure}%[htp]
\centering
\includegraphics[width=0.38\textwidth]{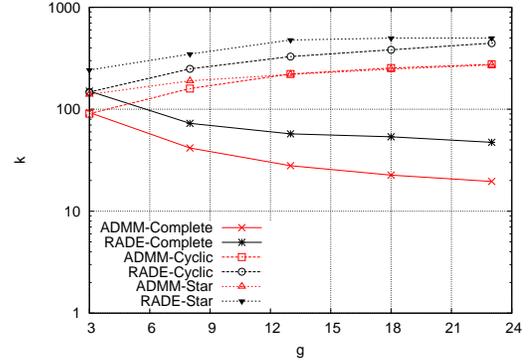}
\caption{Number of time steps until convergence of \ac{rade} when compared to \ac{admm} under different topologies.}
\label{fig:convergence_topology}
\end{figure}

\figurename~\ref{fig:convergence_topology} shows the impact of the virtual topology choice on the convergence performance of \ac{rade} when compared to \ac{admm}. If $\vnum$ is small (three to eight), the impact of the virtual topology on convergence of \ac{rade} and \ac{admm} is minimal. This is because the degree of parallelism (number of  sensors active at the same time) is more restricted by the small number of virtual sensors $\vnum$. In such a scenario, it is convenient for the cloud agent to always arrange the virtual sensors in a star topology. However, as $\vnum$ increases, the impact of the virtual topology becomes significant as the degree of parallelism is higher in a complete topology, enabling \ac{rade} to converge much faster as $\vnum$ gets larger. This convergence becomes slower with star and cyclic topologies. This is because in star and cyclic topologies, only few sensors are active at a time, making \ac{rade} and \ac{admm} converge in a number of steps comparable to that of the \ac{admm}'s sequential implementation. In this later scenario, the cloud agent shall arrange the virtual sensors as a complete topology unless the \ac{sla} permits slower convergence.

Moreover, \ac{rade} converges in a higher number of steps when compared to the conventional \ac{admm}. This is because in \ac{admm}, all sensors are active at each time, and a sensor exchanges its updated variables with all of its neighbors, whereas in \ac{rade}, only disjoint sensor pairs are active at a time and variables are updated only between pairs of sensors. Nevertheless, we argue that this loss in speed of convergence for \ac{rade} is marginal when compared to the significant savings in communication overhead.

\begin{figure}%[htp]
\centering
\includegraphics[width=0.38\textwidth]{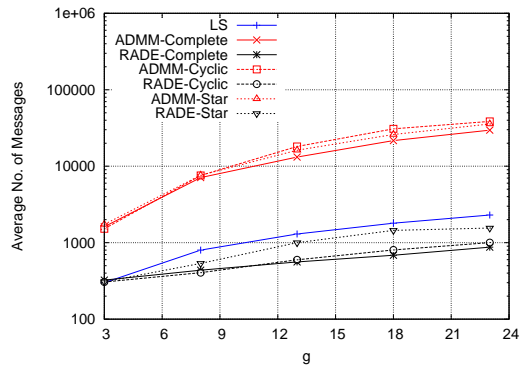}
\caption{Communication overhead when comparing \ac{rade} to \ac{admm} and \ac{ls} under different $\vnum$ values.}
\label{fig:CommOverhead_VN}
\end{figure}		

\figurename~\ref{fig:CommOverhead_VN} shows the total number of $O(\paramsize)$ sized messages exchanged during estimation when comparing \ac{rade}, \ac{admm}, and \ac{ls} for $\obssize = 100$. The number of messages exchanged by \ac{rade} is at least an order of magnitude less than the number of messages generated under \ac{admm}. Also the communication overhead of \ac{rade} is less than the centralized \ac{ls} especially as $\obssize$ becomes large. This savings in communication overhead is attributed to the asynchronous design of \ac{rade} in which messages among sensors are only exchanged if new values of a primal or dual variables are changed away from their specified tolerances.

\section{Conclusion and Discussion}
\label{sec:conclusion}
We propose cloud-based remote sensing algorithms for enabling distributed estimation of unknown parameters via sensor network virtualization. 
The algorithm has the following phases: sensor search, domain pruning, benefit matrix construction, virtual-participatory sensor assignment solver, and distributed estimation. 
Using simulation, we show that the proposed algorithms reduce communication overhead significantly without compromising the estimation error when compared to the traditional ADMM algorithm. We also show that the convergence time of our proposed algorithms maintain linear convergence behavior, as in the case of conventional ADMM.

\bibliographystyle{abbrv}
\bibliography{./references}

\end{document}